\documentclass[oribibl]{llncs}
\usepackage{amsmath}
\usepackage{amsfonts}
\usepackage{amssymb}
\usepackage{graphicx}
\usepackage{epstopdf}
\usepackage{tikz}
\usepackage{algorithm}
\usepackage{algorithmic}
\usepackage{multirow}
\usepackage{url}
\usepackage{relsize}
\usepackage{paralist}

\newcommand{\X}{\boldsymbol{X}}
\newcommand{\Y}{\boldsymbol{Y}}

\newcommand{\W}{\boldsymbol{W}}

\newcommand{\bLambda}{\boldsymbol{\Lambda}}

\newcommand{\R}{\boldsymbol{R}}
\newcommand{\A}{\boldsymbol{A}}

\newcommand{\C}{\boldsymbol{C}}

\newcommand{\J}{\boldsymbol{J}}

\newcommand{\bI}{\boldsymbol{I}}

\newcommand{\M}{\boldsymbol{M}}

\allowdisplaybreaks
\title{\large Generative discriminative models for multivariate inference and statistical mapping in medical imaging}
\author{Erdem Varol, Aristeidis Sotiras, Ke Zeng, Christos Davatzikos}
\institute{Center for Biomedical Image Computing and Analytics \\ University of Pennsylvania \\ Philadelphia, PA 19104}
\begin{document}
\maketitle

\begin{abstract}
This paper presents a general framework for obtaining interpretable multivariate discriminative models that allow efficient statistical inference for neuroimage analysis. The framework, termed generative discriminative machine (GDM), augments discriminative models with a generative regularization term. We demonstrate that the proposed formulation can be optimized in closed form and in dual space, allowing efficient computation for high dimensional neuroimaging datasets. Furthermore, we provide an analytic estimation of the null distribution of the model parameters, which enables efficient statistical inference and p-value computation without the need for permutation testing. We compared the proposed method with both purely generative and discriminative learning methods in two large structural magnetic resonance imaging (sMRI) datasets of Alzheimer's disease (AD) (n=415) and Schizophrenia (n=853). Using the AD dataset, we demonstrated the ability of GDM to robustly handle confounding variations. Using Schizophrenia dataset, we demonstrated the ability of GDM to handle multi-site studies. Taken together, the results underline the potential of the proposed approach for neuroimaging analyses.
\end{abstract}

\section{Introduction}

Voxel-based analysis \cite{ashburner2000voxel} of imaging data has enabled the detailed mapping of regionally specific effects, which are associated with either group differences or continuous non-imaging variables, without the need to define \textit{a priori} regions of interest. This is achieved by adopting a generative model that aims to explain signal variations as a function of categorical or continuous variables of clinical interest. Such a model is easy to interpret. However, it does not fully exploit the available data since it ignores correlations between different brain regions \cite{davatzikos2004voxel}.

Conversely, supervised multivariate pattern analysis methods take advantage of dependencies among image elements. Such methods typically adopt a discriminative setting to derive multivariate patterns that best distinguish the contrasted groups. This results in improved sensitivity and numerous approaches have been proposed to efficiently obtain meaningful multivariate brain patterns \cite{kriegeskorte2006information,sabuncu2012relevance_short, rasmussen2012model_short, cuingnet2013spatial_short,grosenick2013interpretable_short,ganz2015relevant_short}. However, such approaches suffer from certain limitations. Specifically, their high expressive power often results in overfitting due to modeling spurious distracter patterns in the data \cite{haufe2014interpretation}. Confounding variations may thus limit the application of such models in multi-site studies \cite{rao2017predictive_short} that are characterized by significant population or scanner differences, and at the same time hinder the interpretability of the models. This limitation is further emphasized by the lack of analytical techniques to estimate the null distribution of the model parameters, which makes statistical inference costly due to the requirement for permutation tests for most multivariate techniques.

Hybrid generative discriminative models have been proposed to improve the interpretability of discriminative models \cite{mairal2012task,batmanghelich2012generative_short}. However, these models also do not have analytically obtainable null distribution, which makes challenging the assessment of the statistical significance of their model parameters. Last but not least, their solution is often obtained through non-convex optimization schemes, which reduces reproducibility and out-of-sample prediction performance. 

To tackle the aforementioned challenges, we propose a novel framework termed \textit{generative-discriminative machine} (GDM), which aims to obtain a multivariate model that is both accurate in prediction and whose parameters are interpretable. GDM combines ridge regression\cite{hoerl1970ridge} and ordinary least squares (OLS) regression to obtain a model that is both discriminative, while at the same time being able to reconstruct the imaging features using a low-rank approximation that involves the group information. Importantly, the proposed model admits a closed-form solution, which can be attained in dual space, reducing computational cost. The closed form solution of GDM further enables the analytic approximation of its null distribution, which makes statistical inference and p-value computation computationally efficient.

We validated the GDM framework on two large datasets. The first consists of Alzheimer's disease (AD) patients (n=415), while the second comprises Schizophrenia (SCZ) patients (n=853). Using the AD dataset, we demonstrated the robustness of GDM under varying confounding scenarios. Using the SCZ dataset, we effectively demonstrated that GDM can handle multi-site data without overfitting to spurious patterns, while at the same time achieving advantageous discriminative performance.

\vspace*{-2ex}
\section{Method}
\subsubsection{Generative Discriminative Machine:} GDM aims to obtain a hybrid model that can both predict group differences and generate the underlying dataset. This is achieved by integrating a discriminative model (i.e., ridge regression \cite{hoerl1970ridge}) along with a generative model (i.e., ordinary least squares regression (OLS)). Ridge and OLS are chosen because they can readily handle both classification and regression problems, while admitting a closed form solution. 

Let $\X \in \R^{n\times d}$ denote the $n$ by $d$ matrix that contains the $d$ dimensional imaging features of $n$ independent subjects arranged row-wise. Likewise, let $\Y \in \R^{n}$ denote the vector that stores the clinical variables of the corresponding $n$ subjects. GDM aims to relate the imaging features $\X$ with the clinical variables $\Y$ using the parameter vector $\J \in \R^{d}$ by optimizing the following objective:
\begin{align}
\min_{\J} \underbrace{\|\J\|_2^2  + \lambda_1 \| \Y - \X\J\|_2^2}_{\text{ridge discriminator}} + \underbrace{\lambda_2 \| \X^T - \J\Y^T\|_2^2}_{\text{OLS generator}}.
\end{align}

If we now take into account information from k additional covariates (e.g., age, sex or other clinical markers) stored in $\C \in \R^{n\times k}$, we obtain the following GDM objective:
\begin{align}\label{gdm}
\min_{\J,\W_0,\A_0}  \underbrace{\|\J\|_2^2  + \lambda_1 \| \Y - \X\J - \C\W_0\|_2^2}_{\text{ridge discriminator}} + \underbrace{\lambda_2 \| \X^T - \J\Y^T - \A_0\C^T\|_2^2}_{\text{OLS generator}},
\end{align}    
where $\W_0 \in \R^{k}$ contains the bias terms and $\A_0 \in \R^{d\times k}$ the regression coefficients pertaining to their corresponding covariates. The inclusion of the bias terms in the ridge regression term allows us to preserve the direction of the parameter vector that imaging pattern that distinguishes between the groups, while at the same time achieving accurate subject-specific classification by taking into account each sample's demographic and other information. Similarly, the inclusion of additional coefficients in the OLS term allows for reconstructing each sample by additionally taking into account its demographic or other information. Lastly, the hyperparameters $\lambda_1$ and $\lambda_2$ control the trade-off between discriminative and generative models, respectively.

\subsubsection{Closed form solution:}
The formulation in Eq.~\ref{gdm} is optimized by the following closed form solution:
\begin{align}
& \J \nonumber = \left[\bI + \lambda_1 (\X^T\X -\X^T\C(\C^T\C)^{-1}\C^T\X) +\lambda_2(\Y^T\Y -\Y^T \C(\C^T\C)^{-1}\C^T \Y)\right]^{-1} \nonumber \\
&\times \left[(\lambda_1 + \lambda_2) (\X^T \Y - \X^T\C(\C^T\C)^{-1}\C^T \Y)\right],
\end{align}
which requires a $d \times d$ matrix inversion that can be costly in neuroimaging settings. To account for that, we solve Eq.~\ref{gdm} in the subject space using the following dual variables $\bLambda \in \R^{n}$:
\begin{align}
\bLambda = \M^{-1}_{[1:n,1:n]}\bigg(
\bI + \frac{\lambda_2 \X\X^T  \C (\C^T\C)^{-1} \C^T  - \lambda_2 \X\X^T}{1+\lambda_2(\Y^T\Y - \Y^T \C (\C^T\C)^{-1} \C^T \Y)}\bigg)\Y,
\end{align}
where $\M$ is the following $n+k \times n+k$ matrix:
\begin{align}
\M = \left[
\begin{matrix}
-\frac{\X\X^T}{1+\lambda_2(\Y^T\Y - \Y^T \C (\C^T\C)^{-1} \C^T \Y)} - \bI/\lambda_1 & \C\\
 \C^T & 0
\end{matrix}
\right].
\end{align}
The dual variables $\bLambda$ can be used to solve $\J$ using the following equation:
\begin{align}
\J &=  \frac{\lambda_2 \X^T \Y - \lambda_2 \X^T \C (\C^T\C)^{-1} \C^T \Y - \X^T \bLambda}{1+\lambda_2 (\Y^T\Y -  \Y^T \C (\C^T\C)^{-1} \C^T \Y)}.
\end{align}

\subsubsection{Analytic approximation of null distribution:}
Using the dual formulation, the GDM parameters $\J$ can be shown to be a linear combination of the group labels $\Y$ and the following matrix $\mathbf{Q}$:
\begin{align}
\mathbf{Q} = \frac{\lambda_2 \X^T - \lambda_2 \X^T \C (\C^T\C)^{-1} \C^T-\X^T \M^{-1}_{[1:n,1:n]}\bigg(
\bI + \frac{\lambda_2 \X\X^T  \C (\C^T\C)^{-1} \C^T  - \lambda_2 \X\X^T}{1+\lambda_2(\Y^T\Y - \Y^T \C (\C^T\C)^{-1} \C^T \Y)}\bigg)}{1+\lambda_2( \Y^T\Y -  \Y^T \C (\C^T\C)^{-1} \C^T \Y)},\nonumber\\
\end{align}
such that $\J = \mathbf{Q}\Y$ where $\mathbf{Q}$ is approximately invariant to permutation operations on $\Y$. Assuming $\Y$ is zero mean, unit variance yields that $\text{E}(J_i) = 0$ and $\text{Var}(J_i) = \sum_j Q_{i,j}^2$ under random permutations of $\Y$~\cite{varol2018midas_short,varol2018regionally_short}. Asymptotically this yields that $J_i \rightarrow \mathcal{N}(0,\sqrt{\sum_j Q_{i,j}^2)}$, which allows efficient statistical inference on the parameter values of $J_i$.
\section{Experimental validation}
We compared GDM with a purely discriminative model, namely ridge regression \cite{hoerl1970ridge}, as well as with its generative counter-part, which was obtained through the procedure outlined by Haufe et al. \cite{haufe2014interpretation}. We chose these methods because their simple form allows the computation of their null distribution, which in turns enables the comparison of the statistical significance of their parameter maps. The covariates (i.e. $\C = [\text{age}~\text{sex}]$) were linearly residualized using the training set for ridge regression and its generative counterpart.

We used two large datasets in two different settings. First, we used a subset of the ADNI study, consisting of 228 controls (CN) and 187 Alzheimer's disease (AD) patients, to evaluate out-of-sample prediction accuracy and reproducibility. Second, we used data from a multi-site Schizophrenia study, which consisted of 401 patients (SCZ) and 452 controls (CN) spanning three sites (USA n=236, China n=286, and Germany n=331), to evaluate the cross-site prediction and reproducibility of each method.

For all datasets, T1-weighted MRI volumetric scans were obtained at 1.5 Tesla. The images were pre-processed through a pipeline consisting of (1) skull-stripping; (2) N3 bias correction; and (3) deformable mapping to a standardized template space. Following these steps, a low-level representation of the tissue volumes was extracted by automatically partitioning the MRI volumes of all participants into 151 volumetric regions of interest (ROI). The ROI segmentation was performed by applying a multi-atlas label fusion method. The derived ROIs were used as the input features for all methods.

\subsubsection*{Analytical approximation of p-values}
\begin{figure}[!htb]
\centering
\includegraphics[width=0.3\linewidth]{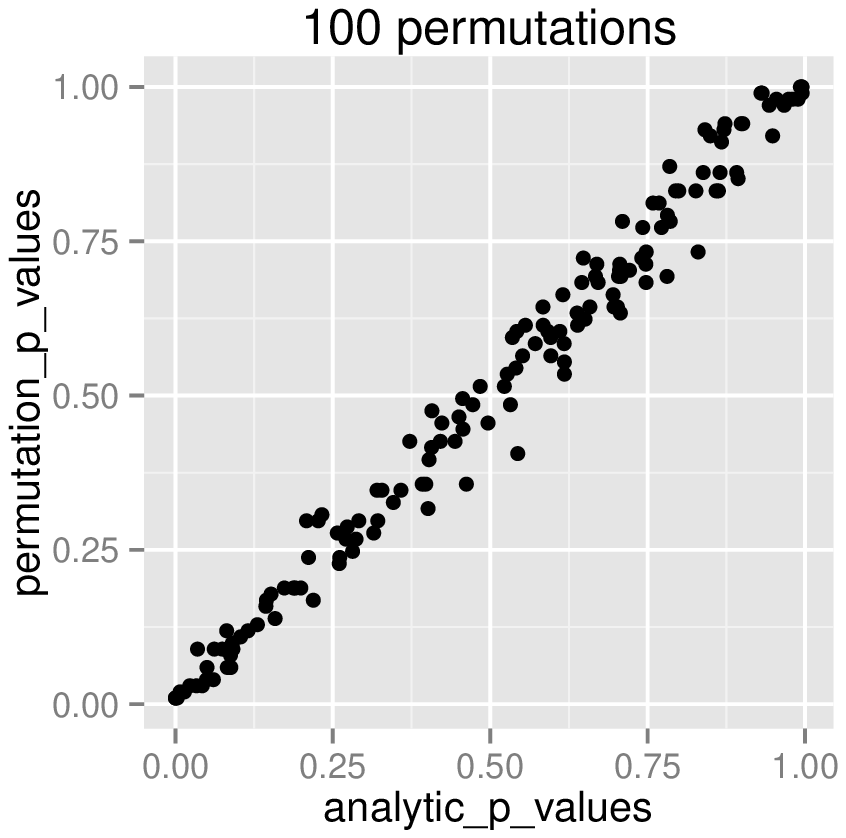}\includegraphics[width=0.3\linewidth]{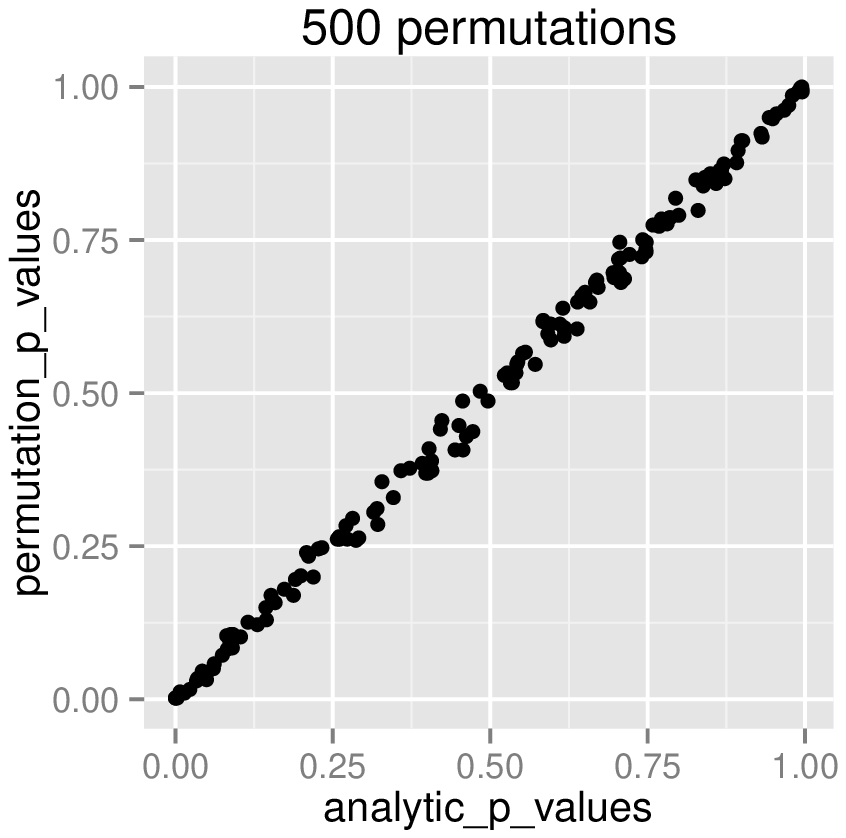}\includegraphics[width=0.3\linewidth]{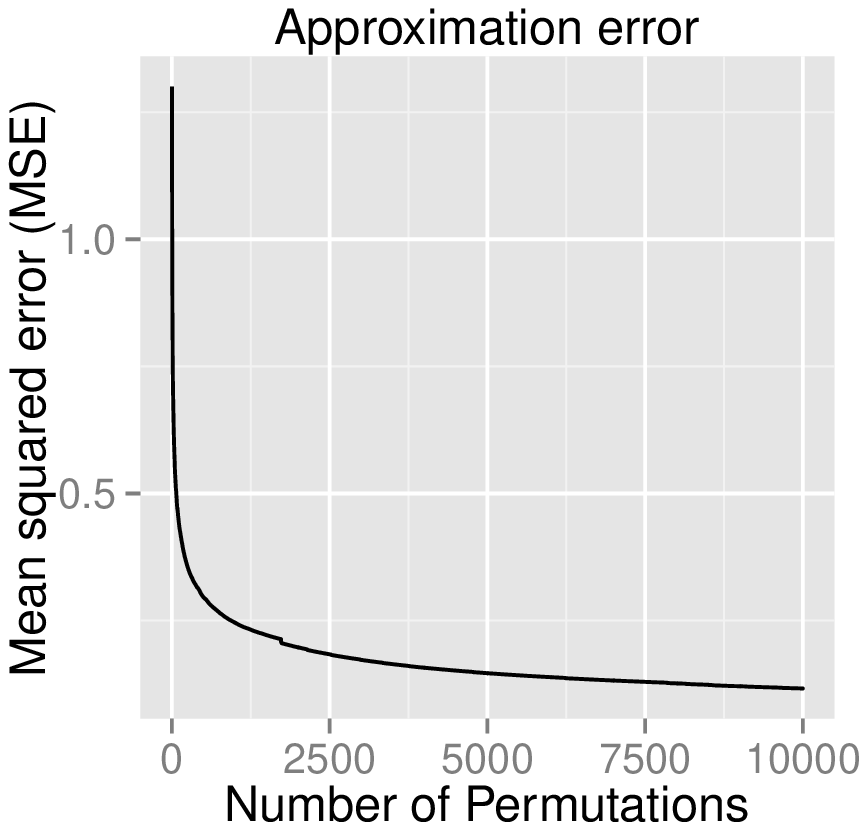}
\caption{Comparison of permutation based p-values of GDM with their analytic approximations at varying permutation levels.}\label{pvals}
\end{figure}

To confirm that the analytical approximation of null distribution of GDM is correct, we estimated the p-values through the approximation technique as well as through permutation testing. A range of 10 to 10,000 permutations was applied to observe the error rate. This experiment was performed on the ADNI dataset. The results displayed in figure~\ref{pvals} demonstrate that the analytic approximation holds with approximately $O(1/\sqrt{\# \text{permutations}})$ error.

\subsubsection*{Out-of-sample prediction and reproducibility} To assess the discriminative performance and reproducibility of the compared methods under varying confounding scenarios, we used the ADNI dataset. We simulated four distinct training scenarios in increasing potential for confounding effects: $\bullet$ Case 1: $50\%$ AD +  $50\%$ CN subjects, mean age balanced, $\bullet$ Case 2: $75\%$ CN + $25\%$ AD, mean age balanced, $\bullet$ Case 3: $50\%$ AD +  $50\%$ CN, oldest ADs, youngest CNs, $\bullet$ Case 4: $75\%$ CN + $25\%$ AD, oldest ADs, youngest CNs. 

All models had their respective parameters ($\lambda_1,\lambda_2 \in \lbrace 10^{-5},\ldots,10^2 \rbrace$) cross-validated in an inner fold before performing out-of-sample prediction on a left out test set consisting of equal numbers of AD and CN subjects with balanced mean age. Furthermore, the inner product of training model parameters was compared between folds to assess the reproducibility of models. Training and testing folds were shuffled 100 times to yield a distribution.

\begin{figure}[!htb]
\centering
\includegraphics[width=0.24\linewidth]{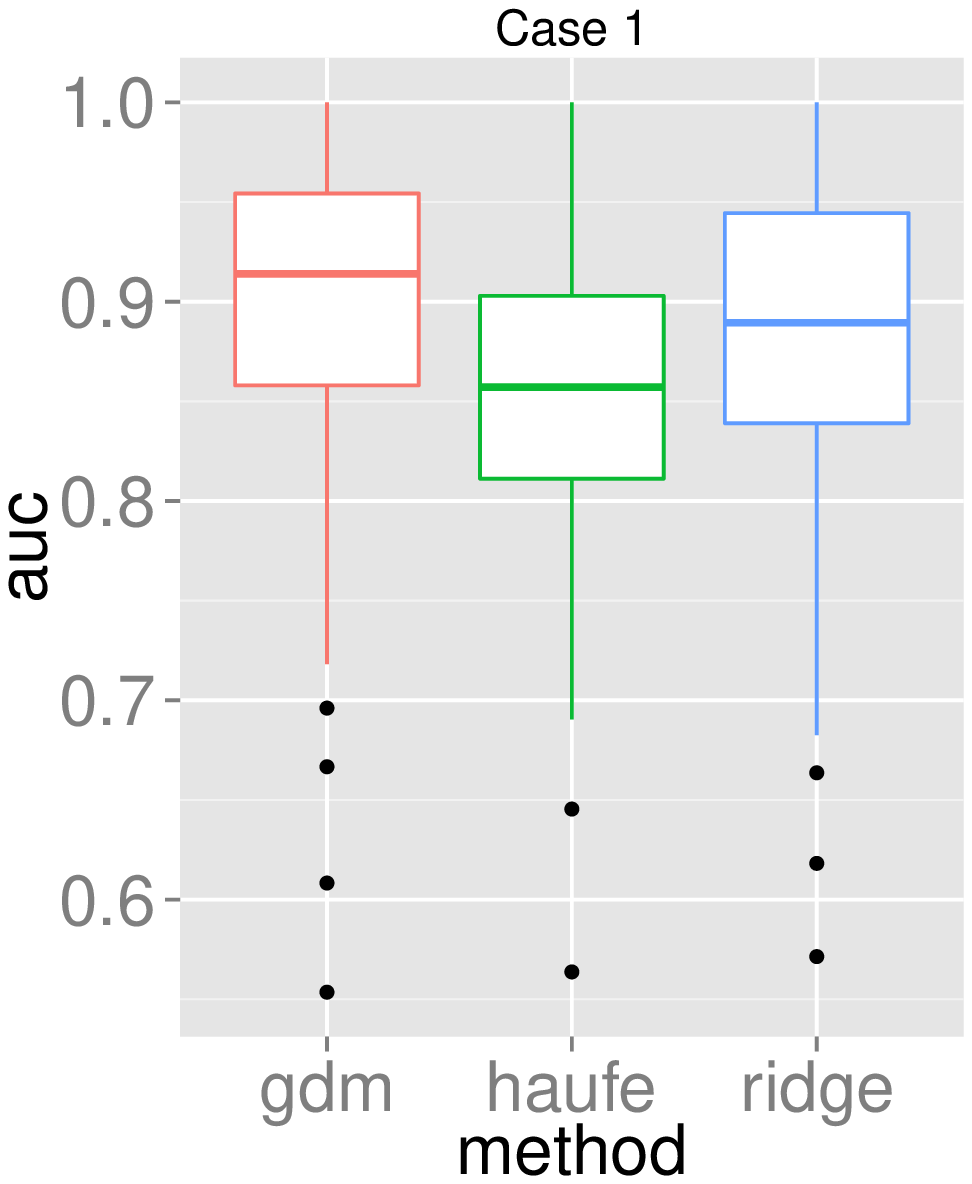} \includegraphics[width=0.24\linewidth]{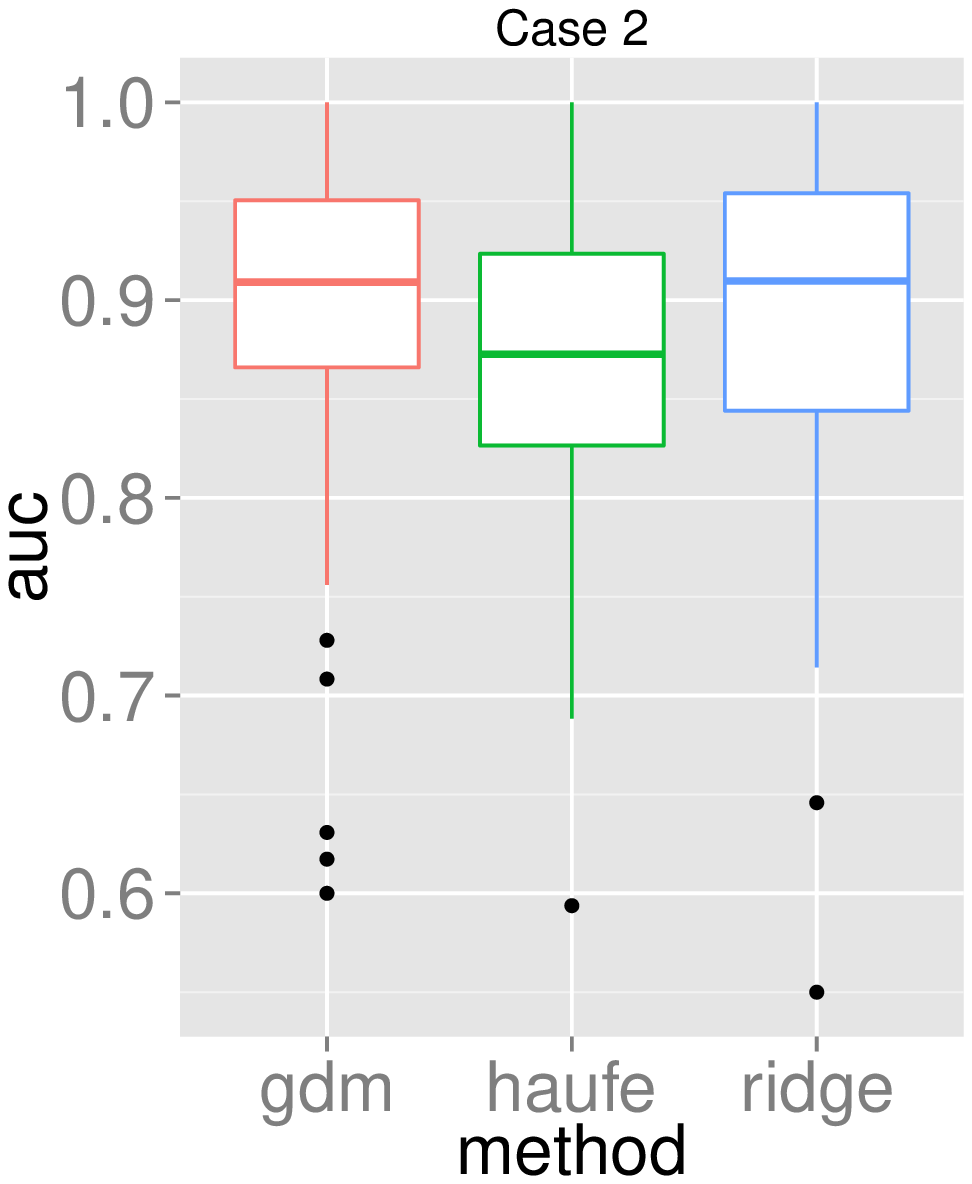} \includegraphics[width=0.24\linewidth]{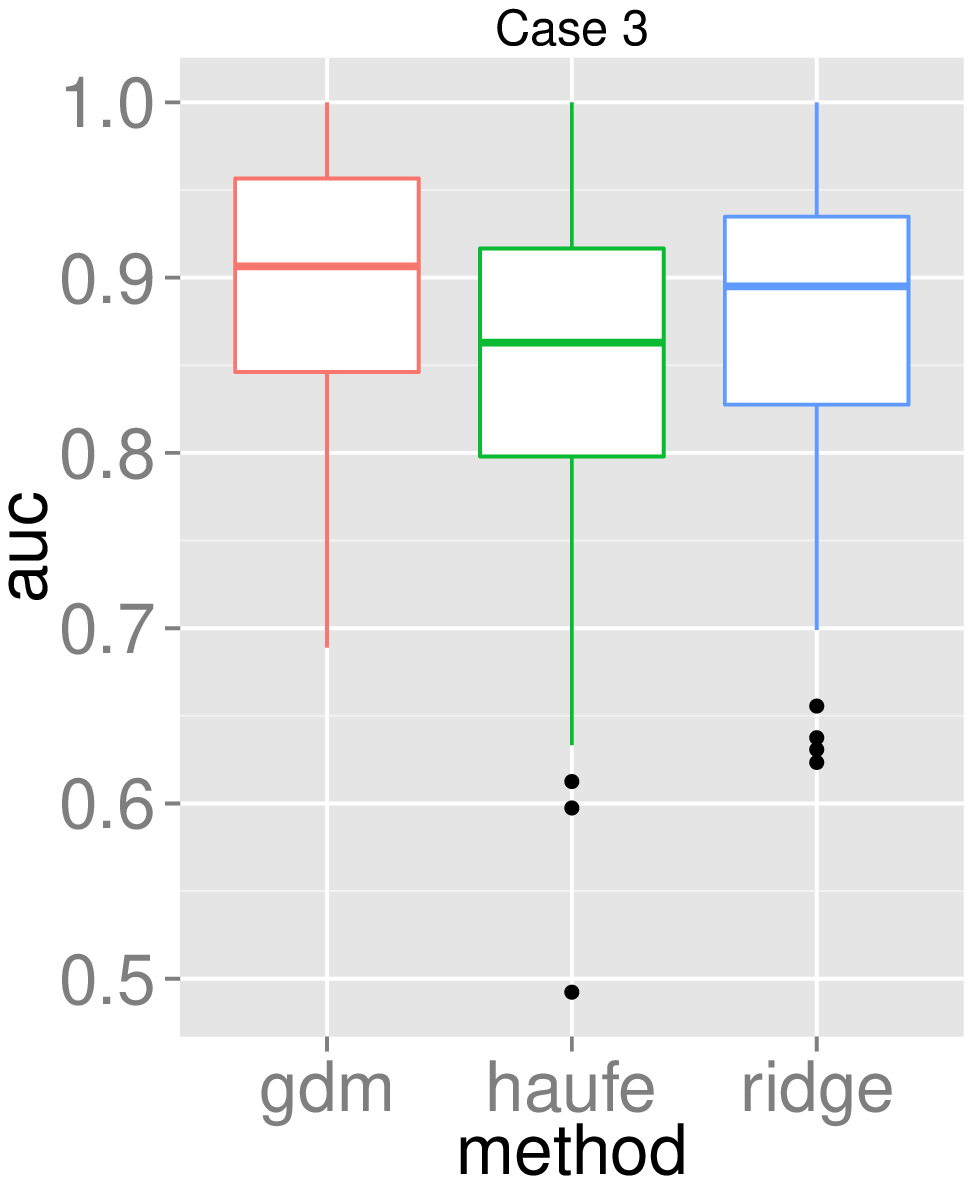} \includegraphics[width=0.24\linewidth]{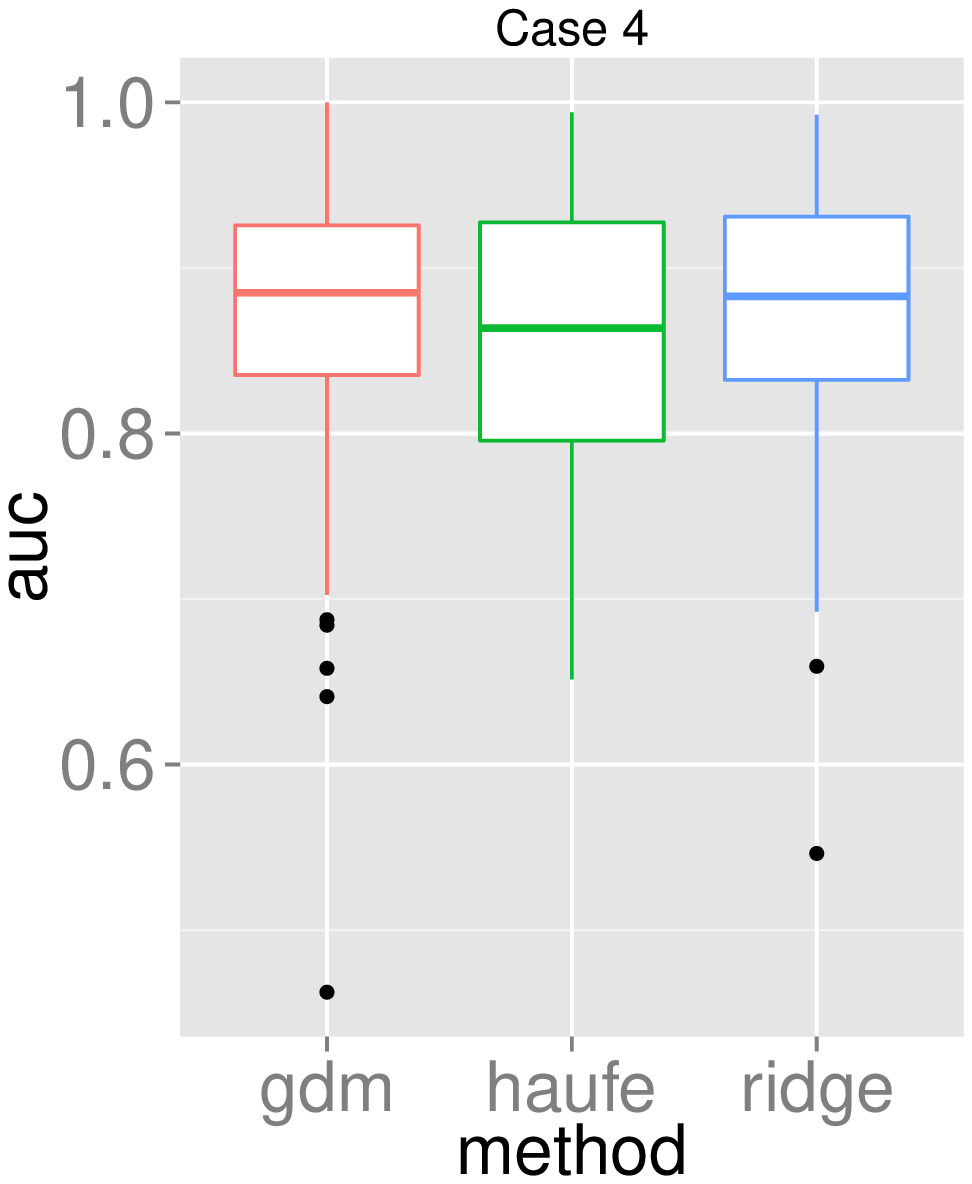}\\
\includegraphics[width=0.24\linewidth]{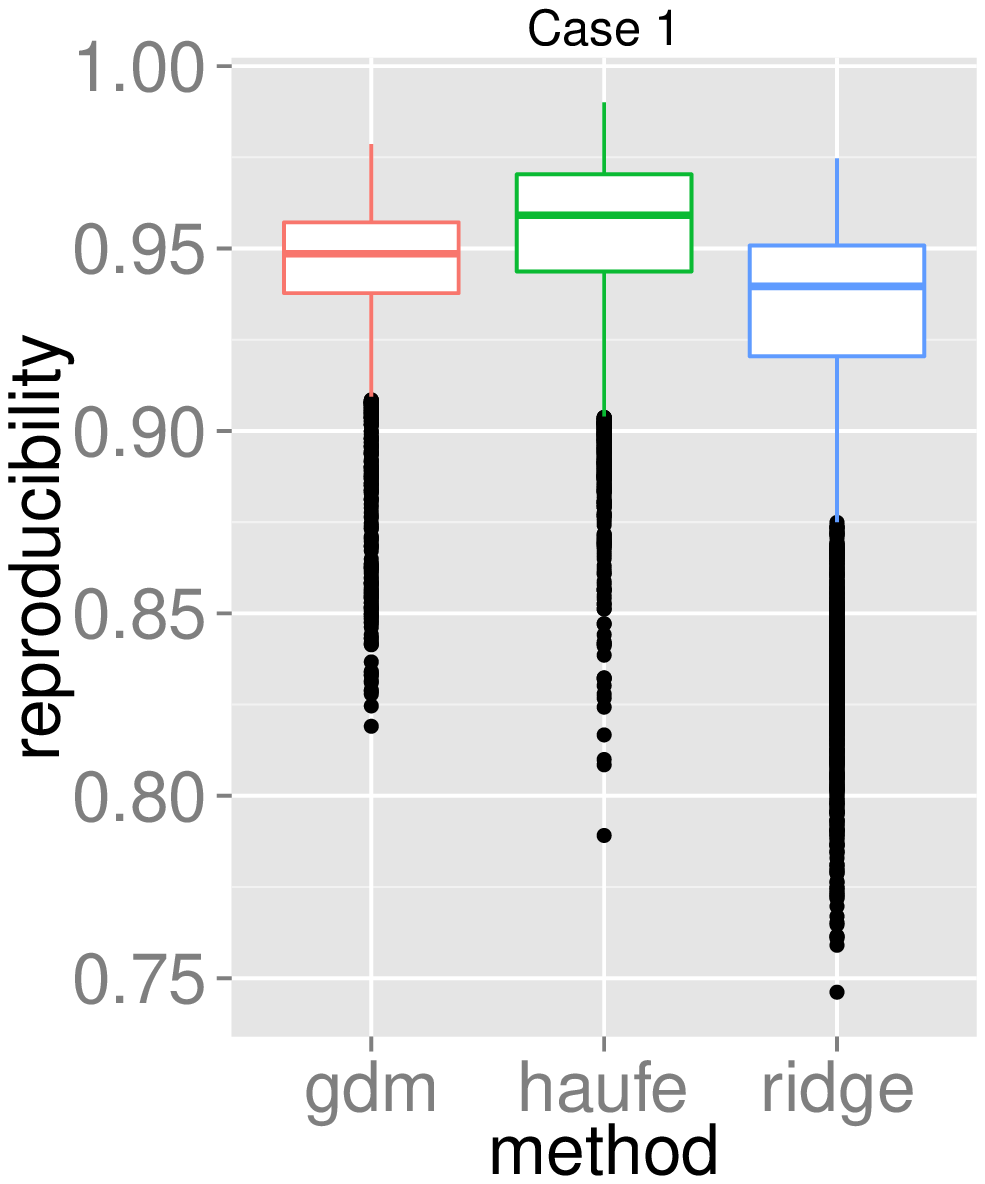} \includegraphics[width=0.24\linewidth]{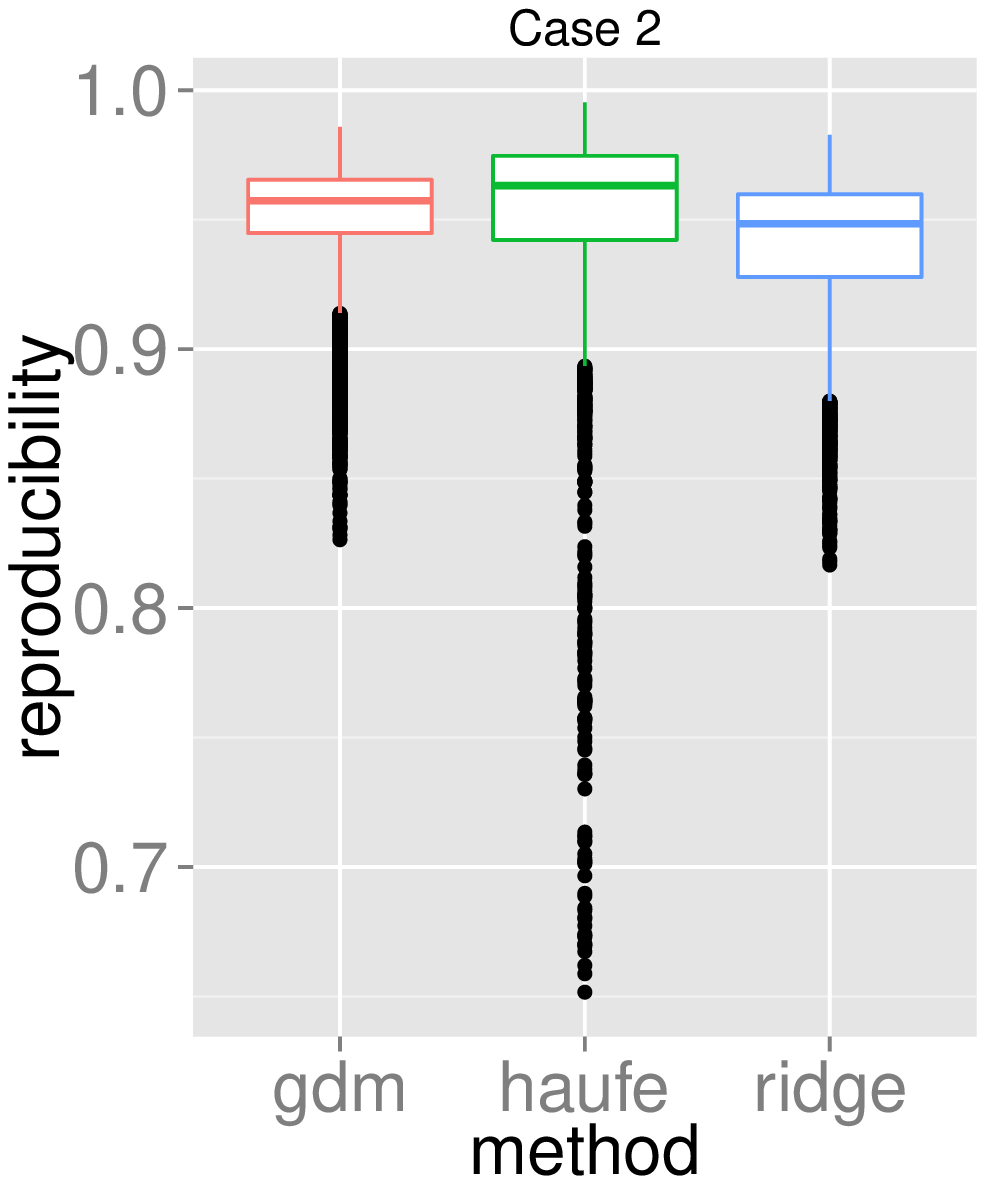} \includegraphics[width=0.24\linewidth]{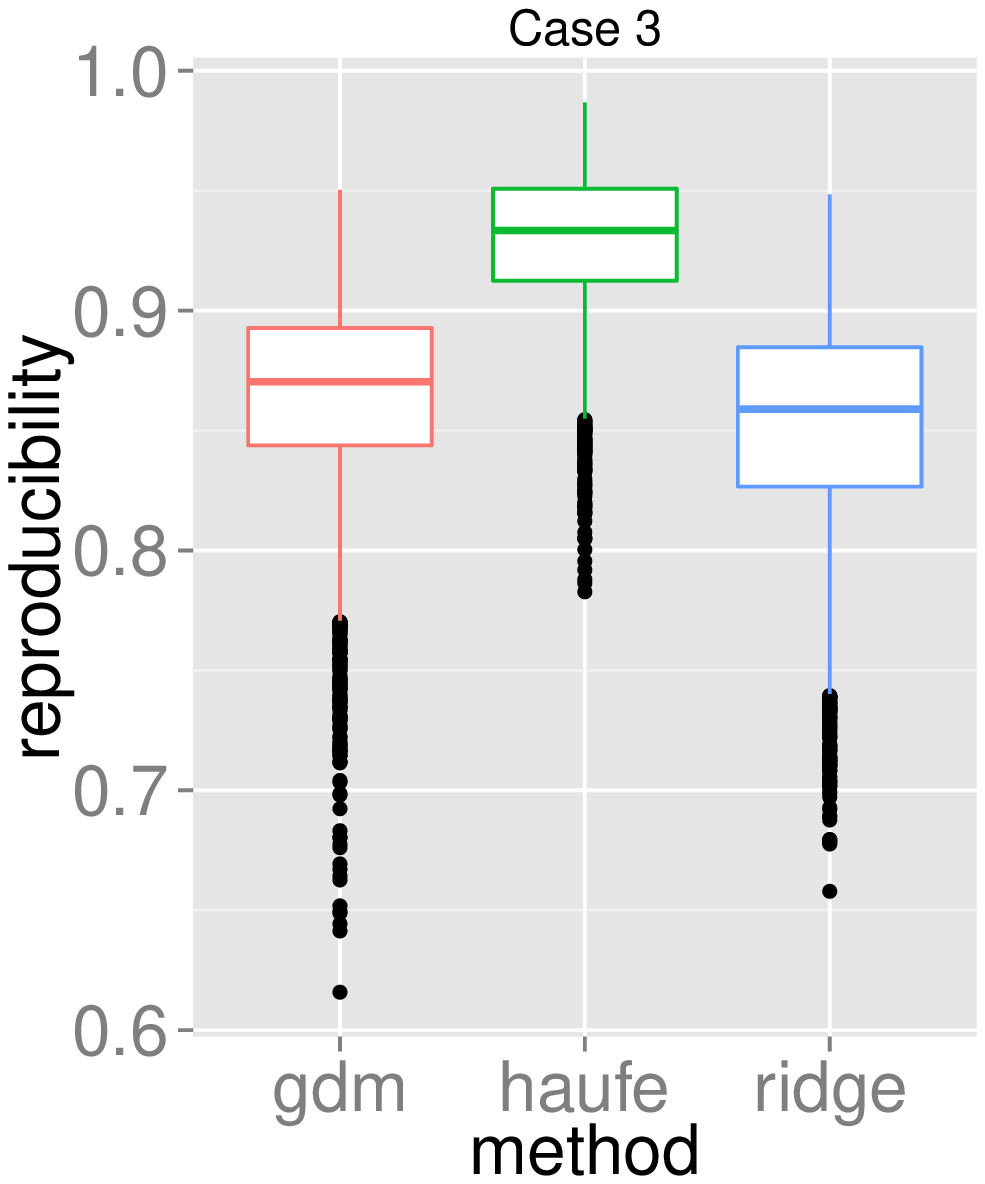} \includegraphics[width=0.24\linewidth]{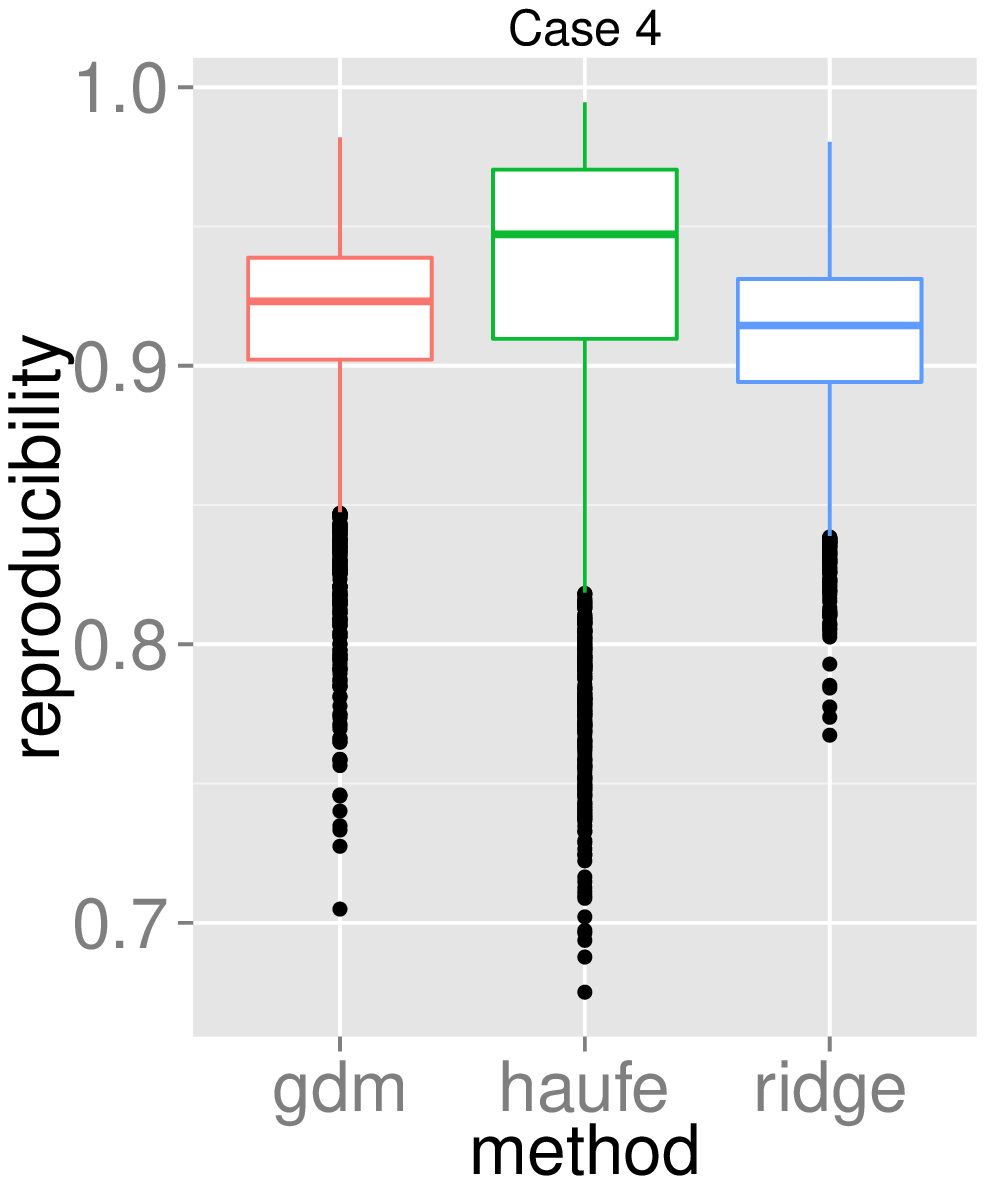}
\caption{Cross validated out-of-sample AD vs. CN prediction accuracies (top row) and normalized inner-product reproducibility of training models (bottom row) for varying training scenarios and all compared methods.}\label{ad_results}
\end{figure}

The prediction accuracies and the model reproducibility for the above cases are shown in figure~\ref{ad_results}. The results demonstrate that while GDM is not a purely discriminative model, its predictions outperformed ridge regression in all four cases. Regarding reproducibility, the Haufe et al. (2014) procedure yielded the most stable models since it yields a purely generative model. However, GDM was more reproducible than ridge regression.

\subsubsection*{Multi-site study}
\begin{figure}[!htb]
\centering
\includegraphics[height=3.6cm]{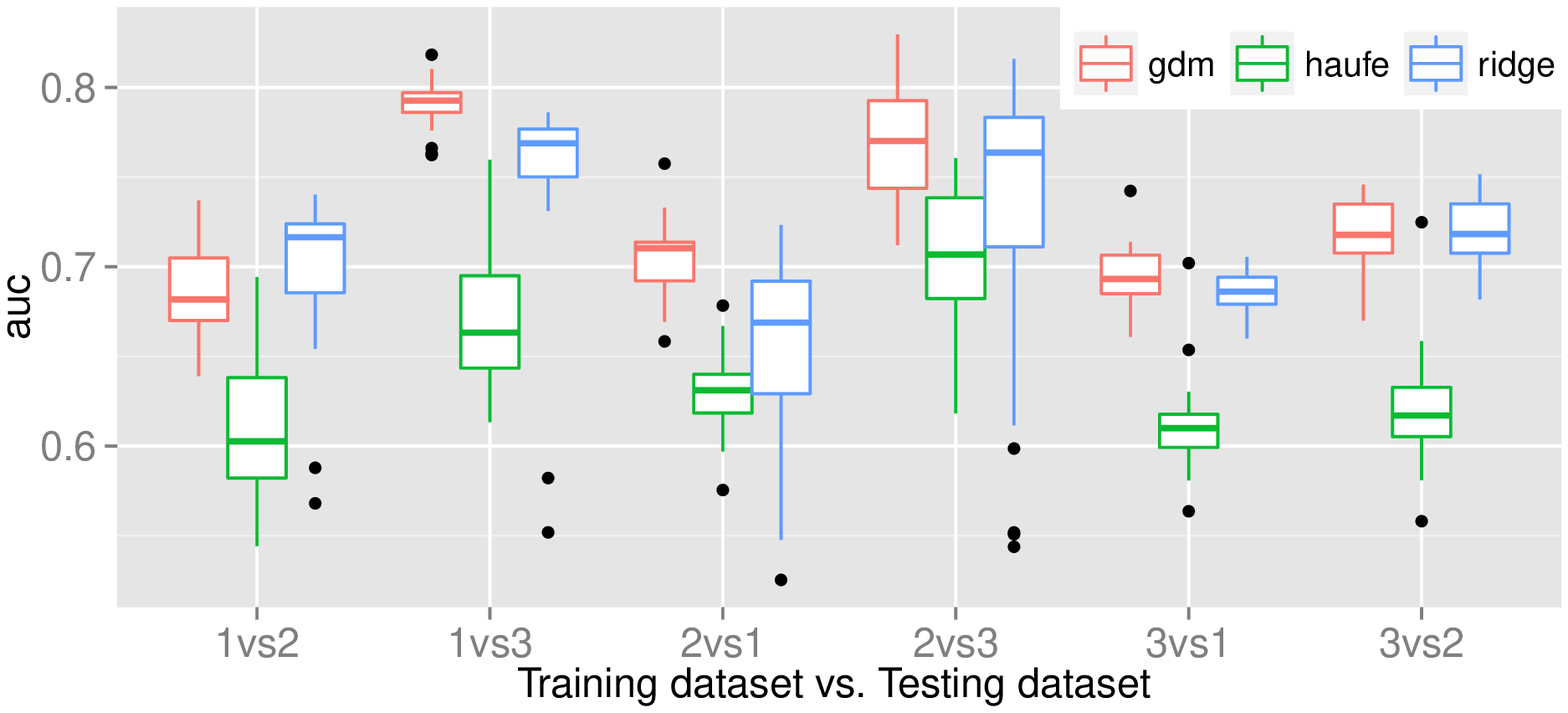}
\includegraphics[height=3.6cm]{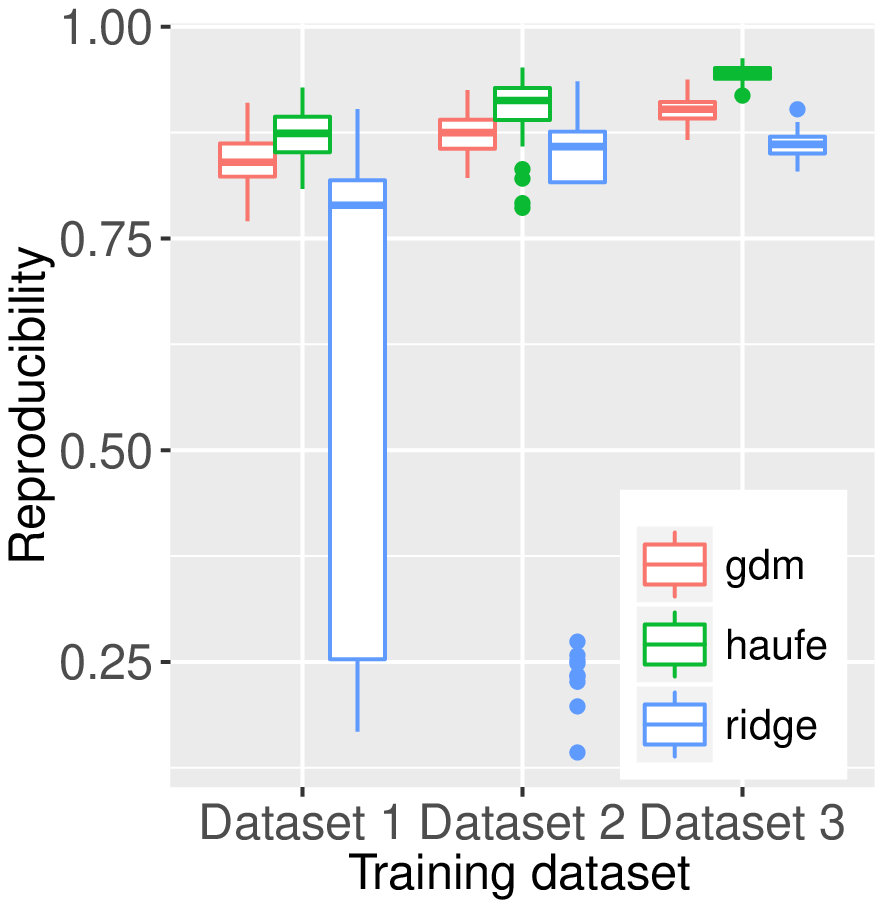}
\caption{Cross validated multi-site SCZ vs. CN prediction accuracies (left) and  normalized inner-product reproducibility of training models (right) for all compared methods.}\label{scz_results}
\end{figure}

To assess the predictive performance of the compared methods in a multi-site setting, we used the Schizophrenia dataset that comprises data from three sites. All models had their respective parameters cross-validated while training in one site before making predictions in the other two sites. Each training involved using $90\%$ of the site samples to allow for resampling the training sets 100 times to yield a distribution. The reproducibility across the resampled sets was measured using the inner product between model parameters. The multi-site prediction and reproducibility results are visualized in figure~\ref{scz_results}.

In five out of six cross-site prediction settings, GDM outperformed all compared methods in terms accuracy. Also, GDM had higher reproducibility than ridge regression, while having slightly lower reproducibility than the generative procedure in Haufe et al. (2014). 

\subsubsection*{Statistical maps and p-values}
\begin{figure}[!htb]
\centering
\includegraphics[width=0.95\linewidth]{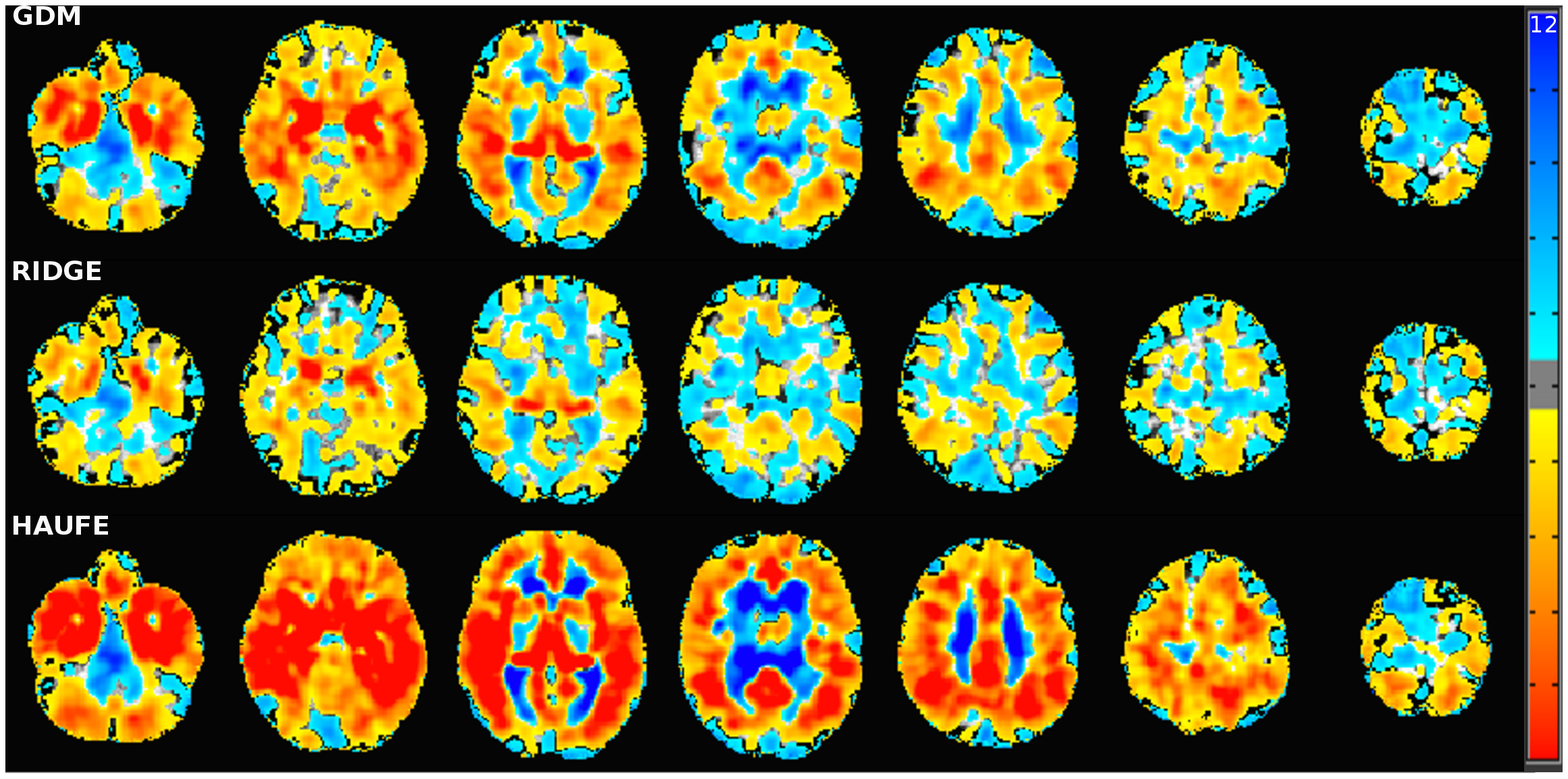}
\includegraphics[width=0.95\linewidth]{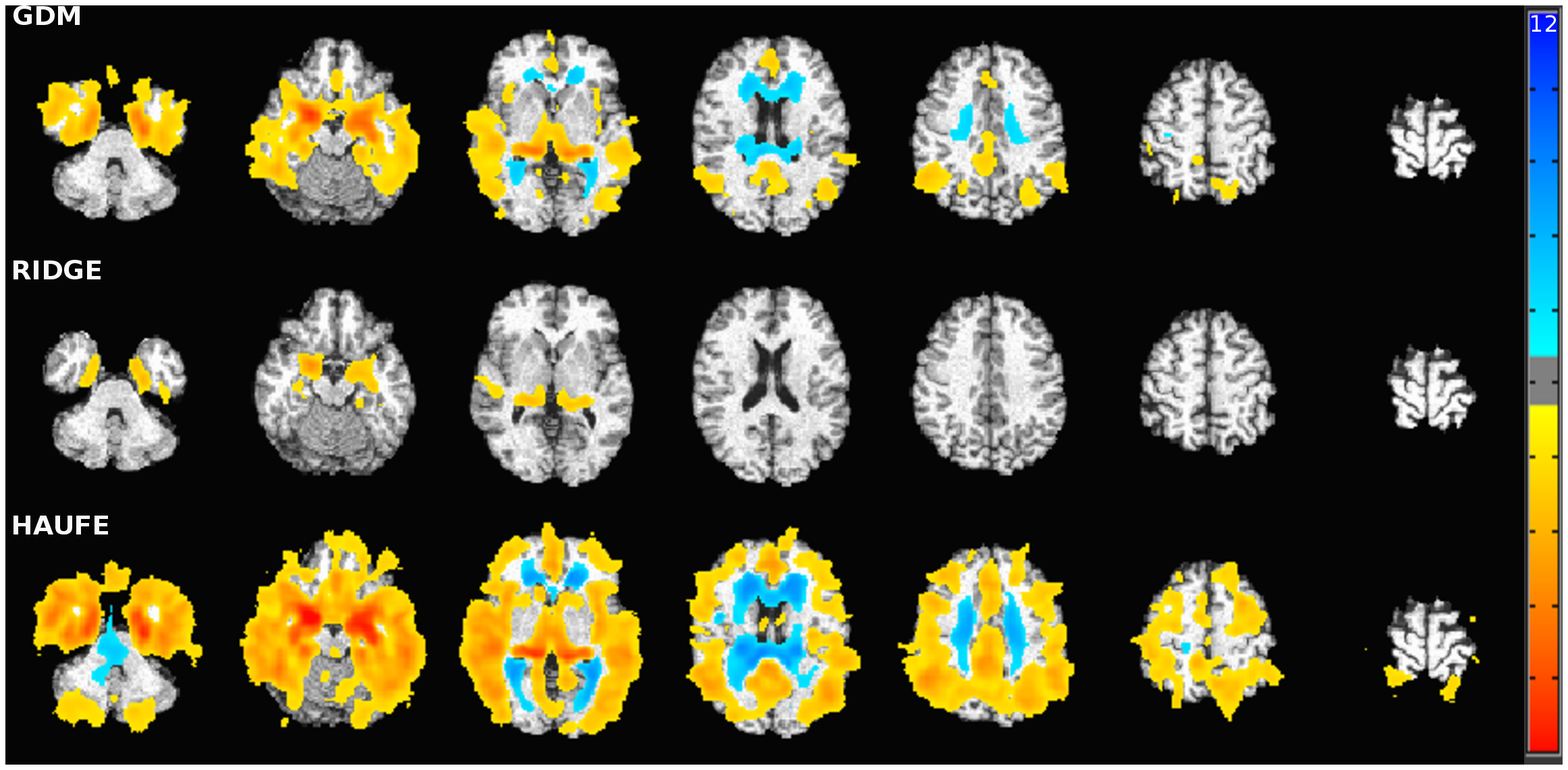}
\caption{Top: Normalized parameter maps of compared methods for discerning group differences between AD patients and controls. Bottom: Parameter log$_{10}$ p-value maps of the compared methods for discerning group differences between AD patients and controls after FDR correction at level $q<0.05$. Warmer colors indicate decreasing volume with AD, while colder colors indicate increasing volume with AD.}\label{p_maps}
\end{figure}

To qualitatively assess and explain the predictive performance of the compared methods for the AD vs. CN scenario, we computed the model parameter maps using full resolution gray matter tissue density maps for the ADNI dataset (Fig.~\ref{p_maps} top). Furthermore, since the null distribution of GDM, as well as ridge regression, can be estimated analytically, we computed p-values for the model parameters and displayed the regions surviving false discovery rate (FDR) correction
\cite{benjamini1995controlling} at level $q<0.05$ (Fig.~\ref{p_maps} bottom).

The statistical maps demonstrated that both GDM and Haufe procedure yield patterns that accurately delineate the regions associated with AD, namely the widespread atrophy present in the temporal lobe, amygdala, and hippocampus. This is in contrast with the patterns found in ridge regression that resemble a hard to interpret speckle pattern with meaningful weights only on hippocampus. This once again confirmed the tendency of purely discriminative models to capture spurious patterns. Furthermore, the p-value maps of the Haufe method and ridge regression demonstrate the wide difference between features selected by generative and discriminative methods and how GDM strikes a balance between the two to achieve superior predictive performance.
\vspace*{-2ex}
\vspace*{-6ex}

\section{Discussion $\&$ Conclusion}
\vspace*{-1ex}
The interpretable patterns captured by GDM coupled with its ability to outperform discriminative models in terms of prediction underline its potential for neuroimaging analysis. We demonstrated that GDM may obtain highly reproducible models through generative modeling, thus avoiding overfitting that is commonly observed in neuroimaging settings. Overfitting is especially evident in multi-site situations, where discriminative models might subtly model spurious dataset effects which might compromise prediction accuracy in an out-of-site setting. Furthermore, by using a formulation that yields a closed form solution, we additionally demonstrated that is possible to efficiently assess the statistical significance of the model parameters.

While the methodology presented herein is analogous to generatively regularizing ridge regression with ordinary least squares regression, the framework proposed can be generalized to include generative regularization in other commonly used discriminative learning methods. Namely, it is possible to augment linear discriminant analysis (LDA), support vector machine (SVM), artificial neural network (ANN) objective with a similar generative term to yield an alternative generative discriminative model of learning. However, the latter two cases would not permit a closed form solution, making it impossible to analytically estimate a null distribution.
\vspace*{-2ex}

\bibliographystyle{splncs03}
\bibliography{refs_full}

\begin{thebibliography}{10}
\providecommand{\url}[1]{\texttt{#1}}
\providecommand{\urlprefix}{URL }

\bibitem{ashburner2000voxel}
Ashburner, J., Friston, K.J.: Voxel-based morphometry-the methods. Neuroimage
  11(6),  805--821 (2000)

\bibitem{batmanghelich2012generative_short}
Batmanghelich, N.K., et~al.: Generative-discriminative basis learning for
  medical imaging. IEEE transactions on medical imaging  31(1),  51--69 (2012)

\bibitem{benjamini1995controlling}
Benjamini, Y., Hochberg, Y.: Controlling the false discovery rate: a practical
  and powerful approach to multiple testing. J. of the royal stat. society.
  (1995)

\bibitem{cuingnet2013spatial_short}
Cuingnet, R., et~al.: Spatial and anatomical regularization of svm: a general
  framework for neuroimaging data. IEEE PAMI  35(3),  682--696 (2013)

\bibitem{davatzikos2004voxel}
Davatzikos, C.: Why voxel-based morphometric analysis should be used with great
  caution when characterizing group differences. Neuroimage  23(1),  17--20
  (2004)

\bibitem{ganz2015relevant_short}
Ganz, M., et~al.: Relevant feature set estimation with a knock-out strategy and
  random forests. Neuroimage  122,  131--148 (2015)

\bibitem{grosenick2013interpretable_short}
Grosenick, L., et~al.: Interpretable whole-brain prediction analysis with
  graphnet. NeuroImage  72,  304--321 (2013)

\bibitem{haufe2014interpretation}
Haufe, S., et~al.: On the interpretation of weight vectors of linear models in
  multivariate neuroimaging. Neuroimage  87,  96--110 (2014)

\bibitem{hoerl1970ridge}
Hoerl, A.E., Kennard, R.W.: Ridge regression: Biased estimation for
  nonorthogonal problems. Technometrics  12(1),  55--67 (1970)

\bibitem{kriegeskorte2006information}
Kriegeskorte, N., Goebel, R., Bandettini, P.: Information-based functional
  brain mapping. PNAS  103(10),  3863--3868 (2006)

\bibitem{mairal2012task}
Mairal, J., Bach, F., Ponce, J.: Task-driven dictionary learning. IEEE
  transactions on pattern analysis and machine intelligence  34(4),  791--804
  (2012)

\bibitem{rao2017predictive_short}
Rao, A., et~al.: Predictive modelling using neuroimaging data in the presence
  of confounds. NeuroImage  150,  23--49 (2017)

\bibitem{rasmussen2012model_short}
Rasmussen, P.M., et~al.: Model sparsity and brain pattern interpretation of
  classification models in neuroimaging. Pattern Recognition  45(6),
  2085--2100 (2012)

\bibitem{sabuncu2012relevance_short}
Sabuncu, M.R., Van~Leemput, K.: The relevance voxel machine (rvoxm): a
  self-tuning bayesian model for informative image-based prediction. TMI
  (2012)

\bibitem{varol2018midas_short}
Varol, E., et~al.: Midas: Regionally linear multivariate discriminative
  statistical mapping. NeuroImage  174,  111--126 (2018)

\bibitem{varol2018regionally_short}
Varol, E., et~al.: Regionally discriminative multivariate statistical mapping.
  In: Biomedical Imaging (ISBI 2018), 2018 IEEE 15th International Symposium
  on. pp. 1560--1563. IEEE (2018)

\end{thebibliography}
	\end{document}